\documentclass{article}
\usepackage{epsfig,natbib}

\begin{document}

\pagestyle{plain}

\renewcommand{\baselinestretch}{1.33}

\normalsize

\Large\centerline{\bf The Origin of the Magnetic
Fields of the Universe:}

\Large\centerline{\bf The Plasma Astrophysics of the
Free Energy of the Universe}

%{\small 12/10/00}
\bigskip
\bigskip
\medskip
\large\indent  \large\indent Stirling A. Colgate$^{1}$, Hui
Li$^{1}$,  Vladimir Pariev$^{1,2,3}$
 
\small\indent  $^1$Theoretical Astrophysics, T6, Los Alamos
National Laboratory, Los Alamos, New Mexico 87545

\small\indent $^2$Steward Observatory, University of Arizona, 
933 North Cherry Avenue, Tucson, AZ 85721

\small\indent $^3$P.N.~Lebedev Physical Institute, 
Leninsky Prospect 53,
Moscow 117924, Russia

 \bigskip
 \bigskip
 \normalsize\noindent

The interpretation of Faraday rotation measure maps of
Active Galactic Nuclei (AGNs) within galaxy
clusters has revealed ordered or coherent regions,
$L_{mag}\sim 50-100$ kpc ($\sim 3\times 10^{23}$ cm),
 that are populated with large, $\sim 30 \mu$ G magnetic
fields. The magnetic energy of these coherent regions is
$L_{mag}^{3}(B^{2}/8\pi) \sim 10^{59-60}$ ergs, and
the total magnetic energy over the whole cluster
($\sim 1$ Mpc across) is expected to be even larger.
Understanding the origin and role of these
magnetic fields is a major challenge to plasma
astrophysics.
A sequence of physical processes that are
responsible for the production, redistribution and
dissipation of these magnetic fields is proposed. 
These fields are associated with single AGNs
within the cluster and therefore with all galaxies
during their AGN (Active Galactic Nucleus or Quasar) phase,
simply because only the central supermassive black
holes ($\sim 10^8 M_{\odot}$) formed during the
AGN phase have an accessible energy of formation, $\sim 10^{61}$
ergs, that can account for the magnetic field energy budget.
An  $\alpha-\Omega$ dynamo process has been proposed that
operates in an accretion disk around a black hole. 
The disk rotation naturally provides a large winding
number, $\sim 10^{11}$ turns, sufficient to make both large
gain and large flux. The helicity of the dynamo can be
generated by the differential plume rotation derived from
star-disk collisions. This helicity generation  process has
been demonstrated in the laboratory and the dynamo gain was
simulated numerically. A liquid sodium analog of the
dynamo is being built. Speculations are that the back
reaction of the saturated dynamo will lead to the formation
of a force-free magnetic helix, which will carry the energy
and flux of the dynamo away from the accretion disk  and
redistribute the field within the clusters and galaxy
walls. The  magnetic reconnection of  a small fraction of
this energy logically is the source of the AGN (Active
Galactic Nucleus or Quasar) luminosity, and the remainder
of the field energy  should then dominate the free energy
of the present-day universe.  The reconnection of this
intergalactic field during a Hubble time is the only
sufficient source of energy necessary to produce an
extragalactic cosmic ray energy spectrum as observed in this
galaxy, and at the same time allow this spectrum to   escape to
the galaxy voids faster than the GZK (black body radiation)
loss.

\newpage

\section*{I. INTRODUCTION}

The total energy released by the growth of supermassive
black holes at the center of nearly every galaxy is large
and can be comparable or even larger than that emitted by
stars in the universe.  Recent observations suggest that
radiation from Active Galactic Nuclei (AGNs or quasars)
might account for only $\sim 10\%$ of this energy.$^{1}$ 
Where did the rest of the energy go? We propose that a
major fraction of this energy has been converted into
magnetic energy and stored in the large scale magnetic
fields primarily external to each galaxy, in galaxy
clusters and ``walls''.  We have envisioned a sequence of
key physical processes that describes this energy
flow.$^{2,3}$ The proposed picture starts with the 2:1
density perturbations of the galactic mass Lyman-$\alpha$
clouds emerging by gravitational instability from the
perturbation structure of the early universe.

The collapse
of these clouds forms the galaxies. Furthermore a small, inner
most fraction, $\sim 10^{-3}$,  collapses via an
accretion disk to form a supermassive central black hole. As
the result of an immense dynamo in this accretion disk,
most of the  accessible energy of formation  of
the black hole is transformed to magnetic field, which
subsequently reconnects, producing high energy particles.  This
field and particle energy  ultimately fill the galaxy walls and
the voids of the  structure of the universe. A consistent
explanation of this filling  process requires a unique
process, apparently analogous to the "flux conversion" of
the spheromak and reverse field pinch. Hence plasma physics
is central to the understanding of the flow of free
energy and even structure of the universe.

Here, we present some of our recent results in
understanding this sequence of phenomena.  In
Section II we summarize the observational evidence for the
existence of immense magnetic fields, mostly based
on Faraday rotation  of AGNs in galaxy clusters and on
synchrotron luminosity measurements of all radio galaxies. 
We briefly discuss our theoretical work on the formation
of an accretion disk of the necessary mass  and then how
this forms the supermassive black holes in Section III.  In
Section IV we describe a dynamo process in this disk by
including star-disk collisions, from which the necessary
{\it helicity} for the $\alpha - \Omega$ dynamo is created.
We have demonstrated this flow structure in our laboratory
experiments. In Section V, we show results from the
numerical simulations on the existence of a dynamo.
The likely back reaction of this dynamo is discussed in
Section VI, as well as our preliminary calculations on the
formation of the force-free magnetic helix. In Section VII we
consider further the distribution and possible dissipation of
this magnetic energy.

\section*{II. MAGNETIC FLUX AND ENERGY
IN GALAXY CLUSTERS}

Recently, high quality Faraday rotation measure maps of
synchrotron sources (e.g., jets from AGNs) embedded in galaxy
clusters where the distances are known have become
available .$^{4,5,6,7}$ (The Faraday rotation is the
rotation of the plane of polarized radiation, from a
background radio source, by an aligned field and a plasma
electron density.  This electron density is measured by
x-ray emission.) An important quantity that has
received less discussion previously is {\it the magnitude
of the magnetic flux and energy} (but see
Burbidge$^{8}$).  Figure (1) shows the  Faraday rotation map of
the region illuminated by Hydra A, the most luminous AGN in
the Hydra cluster of galaxies (courtesy
of Taylor \& Perley$^{4}$).  The largest single region of
highest field in this map has approximately the following
properties: the size $L \simeq 50$ kpc (1 kilo par sec =
$3 \times 10^{21}$ cm) and
$B \simeq 33 \mu$G, derived on the basis  that the field
is patchy and is tangled on a 4 kpc scale. This leads to a
startling estimate of flux, $\it{F}
\approx BL^2 \simeq 8 \times 10^41 $ G $cm^2$,
and energy, $W = (B^2/8\pi) L^3 \simeq 4 \times  10^{59}$
ergs, assuming that the  field is closed on the scale of the
jet and is confined only to the 50 kpc region. Measurements of
field averaged throughout  the whole cluster$^5$  indicate that it
is magnetized throughout with a mean field of $\simeq 3 \mu$ G out
to  a radius of $\sim 300$ kpc.  Then the implied flux and energy
are correspondingly larger by a factor of
$\sim 4$  respectively. A similar
conclusion can be reached when a larger sample of 
of galaxy clusters are analyzed using the data
presented in Eilek$^{6}$.   In addition this emission and
strong polarization itself has strongly suggested  synchrotron
emission requiring even larger magnetic fields and energy$^7$.  It
is extraordinary that Faraday rotation is seen at all and
furthermore that it is so highly coherent over such large
dimensions as in Fig. (1). The observation of these large
coherent  Faraday rotation regions requires a magnetized polarized
source with no internal Faraday rotation overlaid by a magnetized
screen of uniform electron density and field.  
       To put the above numbers in
perspective, a typical galaxy like ours has a magnetic
flux and energy of $10^{38}$ Gauss cm$^2$ and $4\times
10^{54}$ ergs, respectively. (A cluster typically contains
hundreds to one thousand galaxies.) The magnitude of the
implied fluxes and energies in galaxy clusters are so
large, $\times 10^4$ and $\times 10^{5-6}$ respectively,
compared to these quantities within standard galaxies that
a new source of energy and a different form of dynamo are
required than that due to the motions of the galaxy itself
in order  to explain the origin of these magnetic fields.

 A supermassive black hole  in an
AGN offers an attractive site for the production of these
magnetic fields since the accessible energy of formation
of a supermassive  black hole of $\sim 10^8 M_{\odot}$ is $\sim
10^{61}$ to $10^{62}$ ergs . We now discuss this possibility in
detail.

\section*{III. FORMATION OF ACCRETION DISK VIA ROSSBY
VORTICES}

 The physics of forming a supermassive
black hole at galaxy centers is poorly understood; neither
is the formation of galaxies. It is believed that
Lyman-$\alpha$ clouds are the first large scale, radius $\simeq
300$ kpc ($10^{24}$ cm), non-linear structures formed in the
early universe when the density contrast exceeds 2:1. The
subsequent gravitational collapse of both the dark matter and
the baryonic matter has been a subject of intense research. In
Figure (2) we present a simplified view on how this collapse
might proceed. The dark matter collapse is dissipationless and
has been extensively simulated using numerical N-body
codes.$^{9}$ This is represented by the black curve in
Figure (2). This  collapse is inhibited by the partial
conservation of the virial energy  of every particle
of dark matter.  This   results in only a small interior mass
fraction   at small radii.  The baryonic matter on the other hand,
shown as the blue curve in Figure (2),  collapses differently
because the molecular collision cross section is so large
that the gas can not interpenetrate itself like dark
matter (cloud surface density $\simeq 10^{-4}$ g $cm^{-2}$),
and so  becomes shock heated and cools by radiation. After the
initial (essentially) free-fall collapse with small (conserved)
angular momentum, a rotating spheroid is formed with a partial
Keplerian (rotation) support. Hence the baryonic collapse to
this point depends critically upon the initial angular
momentum of the Lyman-$\alpha$ cloud characterized by a
parameter $\lambda \simeq v_{cloud}/v_{Keplerian,cloud}$ averaged
over the cloud. Both theoretical estimates by Peebles$^{10}$ and
the numerical calculations by Warren {\it et al.}$^{11}$   give
$0.05 <\lambda < 0.07$  with a wide dispersion. Hence, the cloud is only slowly
rotating slowly.  As the cloud collapses this rotation speeds up
because of the conservation of angular momentum, eventually
reaching the condition of partial centrifugal or Keplerian
support. The radius of partial Keplerian support occurs where
one expects the onset of the McClauren spheroid instabilities
or  where 
$\lambda_{spheroid} \simeq 0.4$.  Since
$v_{Keplerian,spheroid} \propto v_{Keplerian,cloud}
(R_{cloud}/R_{spheroid})^{1/2}$,  then $R_{spheroid}
\simeq (\lambda_{cloud}/ \lambda_{spheroid})^2 R_{cloud}$.
This parameter determines the start of a galaxy structure
and in particular determines the constant rotation velocity
distribution (i.e., the "flat" rotation curve
characteristic of all spiral galaxies). This constancy of
rotational velocity implies that the total baryonic mass
$M_{interior}$ inside a certain radius
$R$ scales as $M_{interior} = M_{spheroid}
(R/R_{spheroid})$, the blue curve in Figure (2). In Figure
(2), we have also plotted the baryonic surface density
$\Sigma$ (gm cm$^{-2}$), the green curve. The surface
density is an important quantity since it directly
determines whether the disk is ``thick'' enough to contain
its heat  against radiative
cooling. In other words, only when the disk
exceeds a certain critical thickness $\Sigma_c$ could the
hydrodynamic structures within the disk survive long
enough to play a role in angular momentum transport. Based
on the opacity estimates, this critical surface density is
$\Sigma_{c} \simeq 100$ gm cm$^{-2}$. The green curve in
Figure (2) shows the increase of $\Sigma$ from the initial
Lyman-$\alpha$ cloud value $10^{-4}$ gm cm$^{-2}$ to
eventually $\Sigma_c$ in several baryonic
collapse phases. Two interesting quantities emerge from
this simple scaling:  $\Sigma_{spheroid} \simeq
(0.4/\lambda_{cloud})^4 \Sigma_{cloud} \simeq 0.1$ g
$cm^{-2}$ is at the start of the flat rotation curve so that 
the radius when $\Sigma_{flat} \approx \Sigma_c$
is $\sim 10$ pc and the total baryonic mass inside this
scale is $M_{disk}\simeq \Sigma_c \pi R^2 \simeq \Sigma_c /
\Sigma_{spheroid} \simeq 10^8 M_{\odot}$. We associate
this size to be the accretion disk size and the mass  to
be the mass of the central black hole. One key step in
allowing the formation of the central black hole described
above is to find a mechanism to explain the  radially
outward transport of angular momentum. We have identified
one plausible process, namely the Rossby vortex
mechanism.$^{12, 13}$ We have found a global
nonaxisymmetric instability in thin disks that can excite
large scale vortices in the disk. Such vortices are shown
to transport angular momentum outward efficiently.$^{14}$ A
numerical simulation showing the production of vortices is
shown in Figure (3). A sufficient condition for the
initiation of the Rossby Vortex Instability, RVI, is a
radial pressure gradient where  $\Delta P/P \sim> 0.1$
such as would be created  by the continuing feeding of
mass  as creates the "flat rotation curve".  This mass
distribution where $M_{interior} \propto R $ results in
a thickness $\Sigma \propto 1/ R$, the green curve in
figure (2).  Because of the torque transmitted
by the vortices, all of the disk  mass is then accreted to a
black hole mass of  $\sim 10^8 M_{\odot}$. This mass in
turn defines the accessible free energy available through
the dynamo to produce these immense magnetic fields.

\section*{IV. THE $\alpha-\Omega$ DYNAMO IN THE BLACK
HOLE ACCRETION DISK}

 An accretion disk is an attractive
site for the dynamo since it has Keplerian differential
rotation and nearly $\simeq 10^{11}$ revolutions close to
the black hole during its formation time of $\sim 10^8$
years. The key ingredient in our proposed $\alpha-\Omega$
dynamo is the star-disk collisions that  provide the
helicity generation process for the dynamo.$^{2, 15}$
We envision a physical sequence for the dynamo that is
shown in Figure (4). Fig. ~4A shows an initial quadrupole
magnetic field within a conducting accretion disk. The
Keplerian differential rotation within the conducting
accretion disk wraps the radial component  of the
quadrupole field into a much stronger toroidal field, Fig.
~4B. This is the $\Omega$-deformation. A plume, driven by a
star-disk collision, carries a fraction of this now
multiplied toroidal flux above the surface of the disk,
Fig.~4C.  Figure ~4D shows that as the plume expands into
the near vacuum away from the disk, the plume rotates
differentially, always in
the same direction,  relative to the rotating frame. This
rotation carries and twists (counter-rotates) the loop of
toroidal flux,
$\sim \pi /2$ radians, into the orthogonal, poloidal plane.
This is the $\alpha$-deformation, or {\it helicity} of the
dynamo process.   Reconnection allows this loop of flux to
merge with the original quadrupole flux, thereby
augmenting the initial quadrupole field.  For positive
dynamo gain, the rate of adding these increments of
poloidal flux must exceed the negative quadrupole
resistive decay rate.  These same large-scale plumes
driven by convection in a star with internal differential
rotation should similarly produce a dynamo.  In the right
panel of Figure (4),  we also show the schematic of a
liquid sodium experiment we are developing that simulates
the astrophysical dynamo in the laboratory. In this case
the magnetic Reynolds number of the sodium Couette flow is
$R_{m,Couette} \simeq 130$ and for the plumes
$R_{m,Plumes}\simeq 15$ as compared to much larger values
for the accretion disk. For positive dynamo gain, the rate
of addition of poloidal flux must be greater than its
decay.  It is only because the toroidal multiplication can
be so large (or that $ R_{m,\Omega}$ can be so large) that
the generation of helicity in this
$\alpha - \Omega$ dynamo can be more rare and episodic. 
This is different from the $\alpha^2$ dynamos of  R\"adler
and Gailitis.$^{16,17}$ Key to both the astrophysical
accretion disk dynamo and the laboratory experiment is
this new source of helicity produced coherently on the
large-scale  by plumes driven off axis  in a rotating
frame.  We have demonstrated how this plume
rotates in a water analog flow visualization
experiment.  An expanding plume partially conserves its
own angular momentum so that the change in moment of
inertia causes a differential rotation. Figure (5) shows
this rotation effect.

\section*{V. KINEMATIC DYNAMO SIMULATIONS}

We have performed kinematic  dynamo calculations 
using a  3--D code calculating time evolution of the vector 
potential of the magnetic field   by  a  time
dependent  velocity flow field.   We use  the vector 
potential, because  $\nabla
\cdot \bf{B}$ remains zero at all times and no periodic
calculational "cleaning"  of $\nabla \cdot \bf{B}$
has to be performed.  The boundary condition is
perfectly conducting  so  that the flux through
the boundary must be constant in time.  This allows
for an initial poloidal  bias field, but thereafter
the flux through the boundary must remain constant. 
Therefore all the flux generated by a  dynamo 
must remain within the box.  Since this is not the case for
either the experiment or the astrophysical circumstance, we
must simulate problems with the walls as far removed from 
the region of action as possible.  A non conducting
boundary condition requires the solution of the external
potential field at each time step and is planned for the
future. Suppose that $\phi$ and  ${\bf A}$ are the scalar
potential of the electric field and  vector potential of
the magnetic field, such that ${\bf B}=\nabla\times {\bf
A}$, ${\bf v}$ is the velocity field, $\eta$ is the
magnetic  diffusivity ($\eta\approx 1/\mbox{Rm}$), $c$ is
the speed of light.  We use the following gauge condition
in our code
                   
$$ c\varphi-{\bf v}\cdot{\bf A}+\eta\nabla\cdot{\bf
A}=0\mbox{.} $$ 

Then, one can derive the equation for the evolution of
${\bf A}$

$$ \frac{\partial{\bf A}}{\partial t}=-A^k \frac{\partial
v^k}{\partial x^i}-({\bf v}\nabla) {\bf A}+\eta\nabla^2{\bf
A}+(\nabla\cdot {\bf A}) \nabla\eta\mbox{.} $$

The resulting magnetic field is obtained by taking curl of
${\bf A}$ at the last time step.  The velocity field is
specified to resemble the actual flow field with no $ J
\times B $ force or "back reaction".  We observe convergence for
the growth rate of the dynamo and for the structure of the growing
magnetic field at a grid size of $31\times 61 \times 61$, which
are the number of grid points in radial, azimuthal, and
vertical directions, respectively. We have simulated two
problems,  the laboratory sodium dynamo experiment and  the
astrophysical accretion disk.  For both problems we have
observed dynamo gain either in the astrophysical case
within the effective numerical diffusion limit of the code,
$R_{m} \simeq 200 $, or in the experimental case for values
of $R_{m} \simeq 120$ and plume frequency within the
experimental limits of $\sim 5$ Hz. We show in  Figure (6)
the experimental simulation. The conditions for gain
at  $R_{m}$ as small as in the planned experiment are more
demanding than the disk where
$R_{m}$ is presumed to be $R_{m}>> 100$.

\section*{VI. FORMATION OF A FORCE FREE HELIX}

We now discuss what happens when the dynamo saturates.
By definition,  a dynamo has  positive gain of the magnetic field
in the conversion of mechanical energy into magnetic energy.  Thus
we expect a magnetic field, starting from an arbitrarily small
"seed" field, to  exponentiate until the linear assumptions break
down, due to  back reaction, and saturation occurs.  This back
reaction in the case of any  dynamo  is the ponderomotive force of
the field acting on the hydrodynamic conducting
fluid flows that made the dynamo action in the first place.  For
the  $\alpha - \Omega$ dynamo the two particular flows are the
Keplerian flow of rotation, the $\Omega$ flow, and the
flow producing the helicity, the $\alpha$ flow. 
We suggest that the  $\alpha$ flow is produced by star disk
collisions, where the stars are an original  small mass fraction
of the original Lyman$-\alpha$ cloud,
$\sim 10^{-3}$ as is determined by absorption  spectra of such
clouds.  The resulting plumes with trapped flux, produced by
the shock interaction of the star at Keplerian velocity with the
disk, are less sensitive to the ponderamotive forces of the
magnetic field as compared to other possible sources of
helicity, e.g. turbulence.  Furthermore we expect the plume
pressure to be larger than the pressure in the disk and still
large than the toroidal component of the field necessary to affect
the Keplerian motion of the disk, e.g. an accretion $\alpha \simeq
0.1$. Therefore the back reaction of the dynamo acts on the
Keplerian flow, which is equivalent to removing its angular
momentum by doing work on the field.  Thus despite the relatively
low density of the disk, the back reaction causes all the kinetic
energy of the disk, and thus of accretion, to be converted into
magnetic energy by an "$\Omega$-saturated" dynamo.

 The increase of magnetic
pressure in and above the disk and the strong differential
rotation of poloidal fields connecting  different parts of
the disk can lead to the efficient expansion of the
fields, away from the disk.  A poloidal flux line connected at
two different radii will undergo differential rotation leading to
a large winding number, $N_{turns}\sim 10^{10}$ turns during the
$10^8$ years of formation.  For a given value of $B_{poloidal}$ external
to the disk and consequently a poloidal flux, ${\it F_{poloidal}}
\simeq R^2 B_{poloidal}$  and with the assumption of a conducting
medium, the toroidal flux will be increased to
${\it F_{toroidal}}
\simeq R^2 B_{poloidal} \times N_{turns}$.  Thus for even modest
values of winding, $B_{toroidal} >> B_{poloidal}$ and at minimum
energy where $B_{\phi} \simeq B_{z}$ then 
$B_{z} / B_{R} >> 1$ and the centrifugal force on plasma on the
field lines becomes negligable compared to the vertical component
of gravity.  The matter will fall back along these field lines
similar to solar arcades, but with the gravitational potenial
$\times 10^{4 to 7}$ greater and for kT of the plasma $\sim 100$
ev. The residual plasma is that necessary to carry the current 
supported by a small electric field of charge separation,
$E_{gravity} \leq c^4 m_{p} /(2eGM_{BH})  \simeq 10^{-4}$
v/cm.

   This very large toroidal flux is not likely to be confined and
instead  will expand to a minimum energy state  leading to a
force-free helix. Since the expanding fields are
``stressed/twisted'', they also carry away energy and angular
momentum from the disk (i.e., Poynting flux). Thus, the
gravitational energy is released via accretion and meanwhile is
being carried away by the magnetic fields, i.e., the back-reaction
of the dynamo. The rate at which the magnetic energy is carried
away from the disk in this case of $\Omega$-quenching is $(B^2/8
\pi) (\pi R^2) v_{Keplerian} = (1 M_{\odot} / yr) c^2/6 
\simeq  10^{46}$ ergs/s.  For
$\langle R \rangle \approx 10 \times 2GM_{BH}/c^2 =  2 \times 
10^{14}$ cm, the mean magnetic field is $\langle B \rangle \simeq
2 \times 10^4$ G.  The  current necessary to bound this field is $I
= 5R B  = 1.5 \times 10^{19}$ amperes.  The density of
current carriers required to carry this current is $\sim
3$ electrons cm$^{-3}$ at $c/3$.  All other matter will
fall back towards the black hole. Such a plasma density
and this field corresponds to a  $\beta n_{e} kT/ (B^2/8
\pi) \simeq  10^{-14}$, implying that force-free is a good
description of the physical  condition as the fields
expand away.

                   We have performed  preliminary
calculations to simulate such an expansion process. We
treat the disk as an infinitely conducting and massive
boundary. The poloidal fields connecting different parts
of the disk are established in a conducting medium above
the disk. Then the disk starts to rotate differentially
(i.e., Keplerian). The calculations are done by solving
the Grad-Shafranov equation assuming axisymmetry and in
steady state. We consider the
force-free limit since both gravity and gas pressure are
expected to play very minor roles. We use the winding
number of each field line as the input control parameter,
which should be conserved if there is no magnetic
reconnection. Due to the nature of  Keplerian rotation
($\Omega \propto R^{-3/2}$),  nearly all the turns are
added to the field lines connected to the innermost part
of the accretion disk. These field lines will also expand
axially the furthest. This is  illustrated in Figure
(7).                  

We further speculate that the axial extension of  the
helix, formed by the winding of the innermost footpoints,
could become a sequence of force-free, quasi-static
equilibria of minimum energy, or Taylor states.$^{18}$ 
This, however, remains to be shown.

\section*{VII. DISTRIBUTION AND DISSIPATION OF MAGNETIC
FLUX AND ENERGY}

 There is, however, a critical problem in understanding how
the helix would expand radially, in addition to its axial
expansion. The outer radial boundary of the helix is the
pressure of the conducting interstellar or intergalactic
medium (ISM/IGM). The magnetic pressure is likely to
exceed greatly that of the ambient medium so that we
expect the helix to expand radially  just as it progresses
axially related to the Keplerian speed of its
footpoints.   The radial  expansion of the fields from
the kilogauss level at which they are generated to
the $\mu$ G levels of the IGM  represents a major problem
if the poloidal ($B_{z}$) and toroidal ($B_{\phi}$) fluxes
are conserved separately. Not only will the force-free
minimum energy state, the Taylor state, be destroyed, but
conceptually this expansion takes place at the expense of
$\int P\,dV$ work by the field. If this were to be the
case, then the remaining energy in the field would be
negligible after an expansion in spatial  scale by the
ratio of $\times 10^{10}$  (i.e., going from the size of a
black hole $\sim 10^{14}$ cm to 100 kpc $\sim 3\times
10^{23}$ cm). Consequently, all the above theory of black
hole accretion disk dynamo would not be applicable. 
Fortunately  this problem may have been solved
unexpectedly by the equally unexpected behavior of the
reverse field pinch, RFP, and the magnetic fields in the spheromak 
in the laboratory.   The reverse field is the reversal of the
$B_{z}$ component of a cylindrically symmetric field as
a function of radius. One notes exactly this reversal  of
the $B_{z}$ flux in the helix picture of Fig. (7).  The
experiments show that if the outer boundary is expanded,
or equivalently $B_{\phi}$  flux  is added,  a tearing
mode  reconnection takes place that
generates  $B_{z}$ flux, partially  conserves $B_{\phi}$ 
flux and does this remarkable process with negligible
loss  of magnetic energy.  The  generation of $B_{z}$ flux
at the expense of $B_{\phi}$  energy is called a reverse
field "dynamo", although no mechanical energy is
converted to magnetic energy.  We speculate that this
process allows the force-free helix to distribute its flux
throughout the universe with only a very small loss of
energy. The flux generated by the BH accretion
disk dynamo is $\it{F_{total}} = (\pi R) v_{\phi}
B_{\phi} t \simeq 10^{44}$ G $cm^2$.   Thus we feel that the black
hole accretion disk dynamo can produce and distribute the
necessary flux in  the universe.

                   Finally we note that the small
fractional dissipation, $\sim 5 - 10\%$ expected in the
reconnection of the RFP - helix, fulfills the necessary 
energy source for the AGN phenomena.  The reconnection of
the force free field  is the dissipation of $J_{\parallel}$ by
$J_{\parallel}  E_{\parallel}$. The run-away acceleration of
electrons and ions by $E_{\parallel}$ proceeds until the
acceleration of each is limited by its radiative losses. This
acceleration limiting radiation could produce the
extraordinary spectra of AGNs and quasars.$^{2}$ In conclusion
we note that if even
$10^{-3}$ of this magnetic  energy fills intergalactic
space for each galaxy formed, then the pressure of the
field acting upon the baryonic plasma will affect the
subsequent  adjacent galaxy formation since the energy
released is $\times 10^3$ the virialized precollapse
baryonic energy. Thus, magnetic fields may play an
important role in galaxy formation as well.

\bigskip
\noindent {\bf ACKNOWLEDGEMENTS}
\bigskip

\indent{\indent We are particularly indebted to many
colleagues who have helped and encouraged this work,
particularly John Finn, Burt Wendroff, Warner Miller, Greg
Willet,  and Marc Herant of LANL, Richard
Lovelace of Cornell,  Howard Beckley, Dave Westpfahl,  Dave
Raymond, Ragnar Farrel,  and James Weatherall,  of New
Mexico Tech.  This work has
been supported by  the DOE, under contract W-7405-ENG-36.}

\newpage

\noindent$^{1}$ D. O. Richstone, E. A. Ajhar, R. Bender
{\it et al.}, Nature {\bf 395}, 14 (1998).

 \noindent$^{2}$S. A. Colgate and H. Li, Astrophys. Space
Sci. {\bf 264}, 357 (1999).

 \noindent$^{3}$ S. A. Colgate and H. Li,  IAUS
 195, ASP Conf. Series 334, eds.  P.C.H. Martens and S.
Tsurta 1999

 \noindent$^{4}$ G. B. Taylor and R. A.
Perley, Astrophys. J. {\bf 416}, 554 (1993).

 \noindent$^{5}$ T.E., Clarke, P.P. Kronberg, , and  H.
B\"{o}hringer, 2000, Astrophys. J. Lett in press  

\noindent$^{6}$ J. A. Eilek, F.N. Owen,  \& Q. Wang,
2000, submitted to ApJ.
 
 \noindent$^{7}$ P. P. Kronberg,  Prog. Phys. {\bf 57},
325 (1994).

 \noindent$^{8}$ G. R. Burbidge,
Astrophys. J. {\bf 124}, 416 (1956).

 \noindent$^{9}$ J. F. Navarro, C. S. Frenk, S. D. M.
White, Astrophys. J. {\bf 490}, 493 (1997).

 \noindent$^{10}$ P. J. E. Peebles,  Astrophys. J. {\bf
155}, 393 (1969).

 \noindent$^{11}$ M. S. Warren, Quinn P. J.,  Salmon J. K.,
Zurek W. H., Astrophys. J. {\bf 399}, 405 (1992).

 \noindent$^{12}$ R. V. E. Lovelace, H. Li, S. A. Colgate,
and A. F. Nelson,  Astrophys. J. {\bf 513}, 805 (1999).

\noindent$^{13}$ H. Li, J. M. Finn, R. V. E. Lovelace, and
S. A. Colgate, Astrophys. J. {\bf 533}, 1023 (2000).

\noindent$^{14}$ Li, H., Colgate, S.A., Wendroff, B., \&
Liska, R. 2000, 
Astrophys. J., in press.

  \noindent$^{15}$ H.F. Beckley, S.A. Colgate, V.D.
Romero,  \& R. Ferrel,  Physics of Fluids, submitted (2000).

\noindent$^{16}$ K.-H. R\"adler, M. Epstein, and M.
Sch\"uler, Studis geoph. et geod. (Prague) {\bf 42}, 1
(1988).

\noindent$^{17}$ A. Gailitis, O. Lielausis. S. Dement'ev,
{\it et al.}, Phys. Rev. Let. {\bf 84},4365 (2000).

 \noindent$^{18}$ J. B. Taylor, Rev. Mod. Phys. {\bf 58},
741 (1986).

\newpage
\large{List of Figures}

%\begin{figure}[ht!]
 %\begin{center}
%\hskip -0.05in \epsfig{file=figure1.eps,width=3in}
%\caption[1]
\noindent 1. {The Faraday rotation measurement, (shown in
color) of Hydra A cluster. The contours indicate the
synchrotron emission intensity. The field derived from the
Faraday rotation  is $33\mu $ G with total energy $\simeq
10^{60}$ ergs.  The minimum energy derived from the synchroton 
luminosity is also $\simeq 10^{60}$ ergs (Courtesy of Taylor \&
Perley 1993).}\\
 %\end{center}
 %\end{figure}

 %\begin{figure}[ht!]
 %\begin{center}
%\hskip -0.05in \epsfig{file=figure2.eps,width=3in} 
%\caption[2]
\noindent 2. {{\it (Left)}A simplified description of interior
mass versus radius during the collapse of a galactic mass
Lyman-$\alpha$ cloud. The evolution of dark matter ($\sim
10^{12}M_{\odot}$) and baryonic matter ($\sim
10^{11}M_{\odot}$) are shown as the black and blue
curves, respectively. The slope of the blue curve corresponds
to $M_{interior} \propto R$, the mass distribution inferred
from the observation of the "flat rotation curve". The green
curve shows how the baryonic surface density varies during the
collapse (Ordinate is indicated on the right). It shows that
this thickness reaches the critical thickness, $\simeq 100$ g
$cm^{-2}$ for the formation of the Rossby vortex instability at
$R = 10$pc where $M_{interior} = 10^8 M_{\odot}$ all of which
collapses to a massive black hole. }\\
% \end{center}
%\end{figure}

 %\begin{figure}[ht!]
 %\begin{center}
 %\hskip -0.05in
%\epsfig{file=figure3.eps,width=3in} 
%\caption[3]
\noindent 3. {{\it (Left)} A 2-D simulation of the formation of
the Rossby vortex instability in a Keplerian disk with an
initial radial pressure gradient of $\Delta P /P = 0.5$. These
vortices transmit angular monentum with both a linear as well
as non linear amplitude.}\\
 %\end{center}
% \end{figure}

% \begin{figure}[ht!]
% \begin{center}
%\hskip -0.05in
%\epsfig{file=figure4.eps,width=3in} 
%\caption[4]
\noindent 4. {{\it (Left)} A schematic drawing of an $\alpha -
\Omega$ dynamo in an accretion disk. The initial quadrupole
poloidal field (panel A) is sheared by the differential
rotation in the disk, developing a strong toroidal component
(panel B). As a star passes through the disk, it shock heats
the matter of the disk and lifts up a fraction of the toroidal
flux and produces an expanding plume (panel C). After the
plume and loop of flux is rotated by $\sim \pi/2$ radians,
reconnection is invoked to merge the new poloidal flux with the
original flux (panel D). {\it (Right)} A liquid sodium dynamo
experiment,  mimicking the accretion disk dynamo. The conducting
fluid between the two cylinders is rotated differentially
(Couette flow), shearing the radial component of an
external quadrupole field into a toroidal field. Plumes are
driven off-axis, resembling the star-disk collisions. The
resistivity of the fluid ensures the reconnection.}\\
 %\end{center}
 %\end{figure}

 %\begin{figure}[ht!]
 %\begin{center}
%\hskip -0.05in \epsfig{file=figure5.eps,width=3in} 
%\caption[5]
\noindent 5. {The plume rotation experiment$^{14}$. {\it
(a)} Viewed from the side: A plume is driven upwards and
expands. Bubbles outline the boundary of the plume. {\it
(b)} Viewed from above in a co-rotating frame:  The rising
plume rotates or twists relative to the frame, creating a
coherent, large scale helicity.  The measurements of the
relative rotation  are compared to theory.}\\
% \end{center}
 %\end{figure}

% \begin{figure}[ht!]
% \begin{center}
% \hskip -0.05in
%\epsfig{file=figure6.eps,width=3in} 
%\caption[6]
\noindent 6. {The kinematic dynamo calculation. {\it (a)} An
initial bias poloidal field with its radial and axial
components. {\it (b)} This field is wrapped up by the
differential rotation, developing a toroidal component. {\it
(c)} The dynamo gain as a function of $1/N$, where $N$ is
number of revolutions during which a pair of  plumes are
injected. A marginal positive gain is obtained when a pair of
plumes are injected every 4 revolutions. {\it (d)} The
exponentiating field energy for $N=3$. The pulsed increase
in field energy is due to the  injection of each pair of 
plumes.}\\
% \end{center}
% \end{figure}

%\begin{figure}[ht!]
 %\begin{center}
 %\hskip -0.05in
%\epsfig{file=figure7.eps,width=3in} 
%\caption[7]
\noindent 7. {The helix generated by the dynamo in the disk.
The foot prints of the external quadrupole field lines, {\it
(a)}, generated by the dynamo, are attached to the surface of
the disk {\it (b)}. In the conducting low density plasma,
$\beta << 1$, these field lines are wrapped into a force free 
helix.  This helix extends from the disk carrying flux and
energy, a Poynting flux, calculated with the Grad Shafranoff
equation {\it (c)}.}\\
% \end{center}
 %\end{figure}

 \end{document}